\documentclass[useAMS,usenatbib]{mnras}

\usepackage{psfrag,amsmath,amssymb,color,epsfig,rotating}

\usepackage{graphicx}
\usepackage{multirow}
\usepackage{tabularx} 
\def\HI{{\rm HI}\,} 
\def\k{{\mathbf{k}}} 
\def\x{{\mathbf{x}}} 
\def\k{{\mathbf{k}}}
 
\def\k{{\bf k}} 

\begin{document}

\title [\HI bias]{Modelling the post-reionization neutral Hydrogen (\HI) bias}

\author[Sarkar et al.]{Debanjan
  Sarkar$^{1}$\thanks{debanjan@cts.iitkgp.ernet.in},Somnath Bharadwaj$^{1,
    2}$\thanks{somnath@phy.iitkgp.ernet.in}, Anathpindika
  S.$^{2}$\\ $^1$Centre for Theoretical Studies, Indian
  Institute of Technology Kharagpur, Kharagpur - 721302, India \\ $^2$Department of Physics, Indian Institute of Technology Kharagpur,
  Kharagpur - 721302, India}

\date{} \maketitle

\begin{abstract}
Observations of the neutral Hydrogen (\HI ) 21-cm signal hold the
potential of allowing us to map out  the cosmological large scale
structures (LSS) across the entire post-reionization era ($z \leq
6$). Several experiments are planned to map the
LSS over a large range of redshifts and angular scales, many  of
these  targeting the Baryon Acoustic Oscillations. 
It is important to model the \HI distribution in order to
correctly predict the expected signal, and more so to 
correctly  interpret the results after the signal is detected. 
In this paper we have carried out semi-numerical
simulations to model the \HI distribution and
study the \HI power spectrum $P_{\HI}(k,z)$ across the redshift range $1 \le z
\le 6$. We have modelled the \HI bias as a complex quantity
$\tilde{b}(k,z)$ whose modulus squared $b^2(k,z)$  relates
$P_{\HI}(k,z)$ to the matter power spectrum $P(k,z)$, and whose real part $b_r(k,z)$ quantifies 
the cross-correlation between the \HI and the  matter
distribution. We study the $z$ and $k$ dependence of the bias,  and
present polynomial fits which can be used to predict the bias  across 
$0 \le z \le6$ and $0.01 \le k \le 10 \, {\rm Mpc}^{-1}$. We also  present results for 
the  stochasticity $r=b_r/b$ which is important for cross-correlation studies. 
\end{abstract}


\section{Introduction}
Since its predictions by H. van der Hulst in 1944, the neutral hydrogen (\HI) 21-cm line has become a work horse for observational cosmology.
One of the direct applications of the 21-cm emission is to measure the rotation curve of galaxies \citep[e.g. see][and references therein]{begum-chengalur-karachentsev05}, which is one of the most direct probes of dark matter.
The 21-cm emission is also a very reliable probe of the \HI content of the galaxies for the nearby universe.
Surveys like the HI Parkes All-Sky Survey  \citep[HIPASS;][]{zwaan-meyer05}, the HI Jodrell All-Sky Survey \citep[HIJASS;][]{lang-boyce-kilborn2003}, the Blind Ultra-Deep HI Environmental Survey \citep[BUDHIES;][]{jaffe-poggianti2012} and the Arecibo Fast Legacy ALFA Survey
\citep[ALFALFA;][]{martin-giovanelli12} aim to measure the 21-cm emission from 
individual galaxies at very low redshifts  ($z<<1$) to quantify the \HI distribution in terms of the \HI mass function and the \HI density parameter $\Omega_{HI}$.
These studies also help us in understanding the effects of different environments in which \HI resides.
This method fails at higher redshifts where we cannot identify individual galaxies.
Here the cumulative flux of the 21-cm radiation from high redshift emitters appears as a diffused background radiation.
Measurements of the intensity fluctuations in this diffused background provide us a three dimensional probe of the large scale structures over a large redshift range in the post-reionization era ($z \lesssim 6$) \citep{bharadwaj-nath-sethi01,bharadwaj01,bharadwaj-pandey03,bharadwaj-srikant04}.
The advantage of studying the 21-cm emission in the post-reionization era lies in the fact that the modulation of the signal due to complicated ionizing fields is less and the 21-cm power spectrum is directly proportional to the matter power spectrum which enhances its usefulness as a probe of cosmology \citep{wyithe-loeb09}.
This technique provides an independent estimate of various cosmological parameters \citep{loeb-wyithe08,bharadwaj-sethi-saini09}.
The Baryon Acoustic Oscillations (BAO) are embedded in the power spectrum of 21-cm intensity fluctuations at all redshifts and the comoving scale of BAO can be used as a standard ruler to constrain the evolution of the equation of state for dark energy \citep{wyithe-loeb-geil08,chang08,seo-dodelson10}.

Several existing and upcoming experiments are planned to map this radiation at various redshifts.
A number of methods also have been proposed or implemented to recover the informations from the signal faithfully.
\citet{lah-chengalur-briggs07,lah-pracy-chengalur09} used Giant Metrewave Radio Telescope (GMRT)
observations at $z\sim0.4$ to co-add the 21-cm signals (``stacking") from galaxies with known redshifts 
in order to increase the signal to noise ratio and infer the average \HI mass of the galaxies.
This technique has been extended a little to $z\sim0.8$ by studying the cross-correlation between 21-cm intensity maps and the large scale structures traced by optically selected galaxies to constrain the amplitude of the \HI fluctuations \citep{chang-pen-bandura10,masui-switzer-banavar13}.
\citet{ghosh-bharadwaj-ali-chengalur11b,ghosh-bharadwaj-ali-chengalur11a} devised a method to characterize and subtract the foreground contaminations in order to recover the signal and used $610$ MHz ($z=1.32$) GMRT observations to set an upper limit on the amplitude of the \HI 21-cm signal.
\citet{kanekar-sethi-dwarkanath16} extended the signal stacking technique further to $z \sim 1.3$ using GMRT observations and obtained an upper limit on the average \HI 21-cm flux density.

A number of 21-cm intensity mapping experiments like Baryon Acoustic Oscillation
Broadband and Broad-beam \citep[BAOBAB;][]{pober-parsons13}, BAO from Integrated Neutral Gas
Observations  \citep[BINGO;][]{battye-brown12}, Canadian Hydrogen Intensity Mapping Experiment  \citep[CHIME;][]{bandura14}, the Tianlai project \citep{tianlai-chen12}, Square Kilometre Array 1-MID/SUR \citep[SKA1-MID/SUR;][]{bull-camera15-BAO-with-SKA} have been planned to cover the intermediate redshift range $z \sim 0.5 \,- \, 2.5$ where their primary goal is to measure the scale of BAO, particularly around the onset of acceleration at $z \sim 1$.
Recent studies suggest that observations of 21-cm fluctuations on small
scales, with SKA1, can constrain the sum of the neutrino masses
\citep{villaescusa-navarro-bull-Viel15,pal-guha-sarkar16}.
 Observations with
SKA1-MID can also test  different scalar field dark energy
models \citep{hossain-thakur-guhasarkar-sen16}.
\citet{ali-bharadwaj14} and \citet{bharadwaj-sarkar-ali15} present theoretical
estimates for  intensity mapping at $z \sim 3.35$ with the Ooty Wide Field
Array (OWFA), while \citet{chatterjee16} and
\citet{santos-bull-alonso15}  present similar estimates for the upcoming uGMRT
and SKA2 respectively.

The main observable of the 21-cm intensity mapping experiments is the 21-cm brightness temperature fluctuation power spectrum $P_{T}(k,z)$. This can be interpreted in terms of the \HI power spectrum $P_{\HI}(k,z)$ as, 
\begin{equation}
P_{T}(k,z)=\bar{T}_{\HI}^2(z)P_{\HI}(k,z) \,,
\label{eq:1}
\end{equation}
where 
\begin{equation}
\bar{T}_{\HI}(z)
\simeq 4.0 {\rm mK} (1+z)^2 \left(\frac{\Omega_{gas}(z)}{10^{-3}}\right) \left(\frac{\Omega_{b}h^2}{0.02}\right) \left(\frac{H_0}{H(z)}\right) \,
\label{eq:2}
\end{equation}
is the mean brightness temperature of the \HI 21-cm
emission  \citep{ali-bharadwaj14}.
Here, $\Omega_{gas}(z)$ is the density parameter for the neutral gas which can
be expressed in terms of the \HI density parameter $\Omega_{\HI}(z)$ through a
suitable conversion,  all other symbols have their usual meaning. 
We can interpret the \HI power spectrum $P_{\HI}(k,z)$ in terms of the 
matter power spectrum $P(k,z)$ as
\begin{equation}
P_{\HI}(k,z)=b^2(k,z)P(k,z)\,,
\label{eq:3}
\end{equation}
under the assumption that the \HI traces the underlying matter
distribution with a linear bias $b(k,z)$ which 
 quantifies the clustering  of the \HI relative to that of the 
total matter distribution. It is clear that we will  need independent
estimates of both  $\Omega_{\HI}(z)$ and $b(k,z)$ in order to interpret 
the observable $P_{T}(k,z)$ in terms of  the underlying matter power spectrum
$P(k,z)$. Further, the amplitude of the  expected signal  $P_{T}(k,z)$ is very
sensitive to  both  $\Omega_{\HI}(z)$ and $b(k,z)$, and it is crucial to have
prior estimates of these parameters in order to make precise predictions for
the upcoming experiments \citep{hamsa-tirthankar-refregier15} .

Several measurements of $\Omega_{\HI}(z)$ have been carried out both at low and high redshifts.
At low redshifts ($z \sim 1$ and lower)  we have measurements of
$\Omega_{\HI}$ from \HI galaxy  
surveys
\citep{zwaan-meyer05,martin-papastergis-giovanellihaynes10,delhaize-meyer13},   
Damped Lyman-$\alpha$ Systems (DLAs) observations \citep{rao-turnshek06,meiring-tripp-prochaska11} and \HI
stacking  \citep{lah-chengalur-briggs-colless07,rhee-zwaan-briggs13}.  
At high redshifts ($1 < z < 6$), measurements of $\Omega_{\HI}$ come from the
studies of the Damped Lyman-$\alpha$ Systems (DLAs)
\citep{prochaska-wolfe09,noterdaeme-petitjean-carithers12}. 
Earlier observations indicated  the \HI content of the universe to remain
almost constant with $\Omega_{\HI} \sim 10^{-3}$ over the entire redshift
range $z<6$
\citep{lanzetta-wolfe95,storrielomb96,rao-turnshek00,peroux-mcmahon03}.  
However, some recent studies \citep{sanchez-ramirez-ellison-prochaska15} indicate that $\Omega_{\HI}$ evolves significantly from $z\sim2 \,{\rm to}\, z\sim5$,
although the redshift evolution of $\Omega_{\HI}$ is debatable in the
intermediate redshift range, $z=0.1 \, - \, 1.6$
\citep{sanchez-ramirez-ellison-prochaska15}. A
combination of low redshift data with high redshift observations shows that 
$\Omega_{\HI}$ decreases almost by a factor of $4$ between $z=5$ to
$z=0$ \citep{sanchez-ramirez-ellison-prochaska15,crighton-murphy-prochaska15}.

\citet{martin-giovanelli12} have used \HI selected galaxies to estimate the 
\HI bias $b(k)$ at $z\sim 0.06$. Intensity mapping experiments
have measured the product $\Omega_{\HI} \, b \, r$
\citep{chang-pen-bandura10,masui-switzer-banavar13} by studying the
cross-correlation of the \HI intensity with optical surveys ($r$ here is the
cross-correlation coefficient or ``stochasticity'') while  
\citet{switzer-masui-bandura-calin13} have measured  the combination
$\Omega_{\HI} \, b$, all  these measurements being  at $z <1$.  
We do not, at present,  have any observational constraint on  the \HI bias
$b(k,z)$  at redshifts $z>1$. It is therefore important to model 
$b(k,z)$ as an useful input for  the future 21-cm intensity mapping
experiments.

\citet{marin-gnedin10} have developed an analytic framework for calculating
the large scale \HI bias $b(k,z)$ and studying   its redshift evolution using 
a relation between the \HI mass  $M_{\HI}$ and the halo mass $M_h$  motivated 
by observations of the  $z=0$ \HI mass function. Analytic techniques, however 
are limited in   incorporating  the effects of nonlinear clustering.   
In an alternative approach, \citet{bagla10} have proposed a  semi-numerical 
technique which utilizes a prescription to populate \HI in the halos
identified from dark matter only simulations. The same approach has also been
used by \citet{khandai-sethi-dimatteo11} and \citet{tapomoy-mitra-majumder12}  
to study the \HI power spectrum and the related
bias. \citet{navarro-viel-datta-choudhury14} have used  high-resolution
hydrodynamical N-body simulations along with three different prescriptions for
distributing the \HI. 
\citet{seehars-paranjape15} have proposed a semi-numerical model
for simulating  large maps of the  \HI intensity distribution  at $z<1$. 
The analytic and semi-numerical studies carried out
till date are all limited in that each study is restricted to a few discrete
redshifts.
In a recent paper \citet{hamsa-tirthankar-refregier15} have compiled all the
available  results for the  \HI bias  and interpolated the
values to cover the redshift range $z=0 \,- \, 3.4$. Their study is restricted to 
large scales where it is reasonable to consider a constant $k$ independent
bias $b(z)$.  We do not, at present,  have a comprehensive study which uses a
single technique to study the \HI bias over a large 
$z$ and $k$ range.

In this work, we study : \textbf{(i)} the evolution of the \HI power-spectrum
$P_{\HI}(k,z)$ across the redshift range $1 \le z \le 6$ by using N-Body
simulations coupled with the third \HI distribution model of \citet{bagla10},
\textbf{(ii)} the redshift variation of the complex bias  $\tilde{b}(k,z)$
whose modulus squared,  $b^{2}(k,z)$, relates $P_{\HI}(k,z)$ to the matter
power-spectrum $P(k,z)$, and whose real component  $b_r(k,z)$ quantifies
the cross-correlation between the \HI and 
the total matter distribution, and \textbf{(iii)} the spatial(rather, $k$)
dependence of the bias and present  polynomial fits  which can be used to
predict its variation  over a large $z$ and $k$ range. 
We note that the entire analysis of this paper is
restricted to real space {\it i.e.} it does not incorporate redshift space
distortion arising due to the peculiar velocities. We plan to address the
effect of peculiar velocities in future. 
An outline of the paper follows.

In section \ref{sec:simulations}, we briefly describe the method of simulating
the \HI distribution. In section \ref{sec:results}, we present the results of
our simulations. Section \ref{subsec:fitting} contains the details of the
polynomial fitting for the joint $k$ and $z$ dependence of the biases.  
The values of the fitting parameters are tabulated in Appendix \ref{sec:appendix}.
We finally summarize all the findings and discuss a few current results on the
basis of our simulations in section \ref{sec:summary}.

We use the fitting formula of \citet{eisenstein-hu99} for the $\Lambda$-CDM 
transfer function  to generate the initial matter power 
spectrum. The cosmological parameter values used are as given in 
\citet{planck-collaboration13}.

\section{Simulating the \HI distribution}
\label{sec:simulations}

We follow three main steps to simulate the post-reionization \HI 21-cm  signal. 
In the first step we  use a Cosmological N-body code to simulate the matter
distribution at   the desired redshift $z$. Here we have used  a Particle Mesh
(PM)  N-body code  
developed by \citet{bharadwaj-srikant04}. This `gravity only' code treats the
entire matter content as dark matter and ignores  the baryonic physics.  
The simulations use  $[1,072]^3$ particles in a $[2,144]^3$  regular cubic
grid of spacing $0.07 \, {\rm Mpc}$  with a total  simulation volume (comoving) 
of  $[150.08 \,  {\rm Mpc}]^3$. The simulation particles all have
mass   $10^{8} \, M_{\odot}$ each. 
We have used the standard linear $\Lambda$-CDM power spectrum to set the initial
conditions at $z=125$, and the N-body code was used to evolve the particle
positions and velocities to the  redshift $z$ at which we desire to simulate
the \HI signal.  We have considered $z$ values in the interval $\Delta z =0.5$ in
the range  $z=1$ to $z=6$.

In the next step we employ the Friends-of-Friends (FoF) algorithm
\citep{davis85} to identify collapsed halos in the particle distribution
produced as 
output by the N-body simulations. For the FoF algorithm we have used a  linking
length of  $l_f=0.2$ in  units of the mean inter-particle separation, 
and furthermore, we require  a halo to have at least ten particles. 
This sets $10^{9} \, M_{\odot}$ as the minimum halo mass that is resolved 
by  our simulation. We also verify that the mass distribution of halos so
detected are in good agreement with  the theoretical halo mass function
\citep{jenkins-frenk01,sheth-tormen02} in the mass range $10^9 \leq M \leq
10^{13} \,  M_{\odot}$. Our halo mass range is well in keeping with a recent
study \citep{kim-wyithe-baugh-lagos16} which shows that  at $z \ge 0.5$ 
a dark matter halo mass resolution better than $\sim 10^{10} \, h^{-1} \,
M_{\odot}$ is required to predict  21-cm brightness fluctuations that are
well converged.

The observations of quasar (QSO) absorption spectra suggest that the  diffuse
Inter Galactic Medium 
(IGM) is highly ionized  at redshifts $z \le 6$
\citep{becker-fan01,fan-carilli06,fan-strauss06}. This 
redshift range where the hydrogen neutral fraction has a value $x_{\rm
	\HI}<10^{-4}$  is referred to as the post-reionization era.  Here the 
bulk of the \HI resides  within  dense clumps 
(column  density $N_{\HI} \geq 2\times10^{20}{\rm cm^{-2}}$)
which are seen  as the Damped Lyman-$\alpha$ systems (DLAs)  found in the QSO
absorption  spectra  \citep{wolfe05}.  
Observations indicate that the DLAs contain almost $\sim 80 \,\%$ of the total \HI,
\citep{storrie-lomb00,prochaska-herbertfort-wolfe05,zafar-peroux-popping-milliard13}
and they are the source of the \HI 21-cm radiation in the 
post-reionization era. 
While the origin of the high-$z$ DLAs
is still not  very well 
understood, it is found (eg. \cite{haehnelt00}) that it is 
possible to correctly reproduce many of the observed DLAs properties if it is
assumed that the DLAs are associated with galaxies. 
From the cross-correlation study between DLAs and Lyman Break Galaxies (LBGs) at $z\sim 3$,  \citet{cooke-wolfe-gawiser06} showed that the halos with mass in the range $10^{9}<M_h<10^{12}\,M_{\odot}$ can host the DLAs.
In this work we assume that \HI in the post-reionization era is entirely contained  within  dark matter  halos which also host the galaxies. In the third step of our simulation we populate the halos
identified by the FoF algorithm with \HI.  Here we assume that the \HI
mass  $M_{\HI}$ contained within a halo depends only on the halo mass
$M_h$, independent of the environment of the halo.

At the outset, we expect the \HI  mass to increase with the size of the
halo.   However, observations  at low $z$
indicate that we do not expect the large halos, beyond a certain upper
cut-off halo mass $M_{\rm max}$, 
to contain a significant amount of \HI. For example, very little \HI is found
in the  large  galaxies which typically are ellipticals and in  the clusters of
galaxies \citep[eg. see][and references therein]{serra-oosterloo-morganti12}. Further, we also do not expect the very small halos,  beyond a
certain lower cut-off halo mass $M_{\rm min}$, to contain significant \HI mass.  
The amount of gas contained  in  small halos $(M_h <M_{\rm min} )$ is inadequate  
for it to be self shielded against the ionizing radiation.  
Based on these considerations, 
\citet{bagla10} have introduced several schemes for populating  simulated
halos  to  simulate the post-reionization  \HI distribution.
In our  work we have implemented one of the schemes proposed  by
\citet{bagla10} to populate the halos. This uses 
an approximate relation between the virialized halo mass and the circular
velocity as a function of  redshift 
\begin{equation}
  M_h \simeq
  10^{10}\left(\frac{v_{\rm circ}}{60{\rm km/s}}\right)^3
  \left(\frac{1+z}{4}\right)^{-\frac{3}{2}}M_{\odot}  \,.
  \label{eq:5}
\end{equation}
It is assumed that 
the neutral gas in the halos will be able to shield itself from the ionizing
radiation only if the halo's circular velocity exceeds $v_{\rm circ}\sim30$
km/s, which  sets the lower mass limit of the halos $M_{\rm min}$. The
upper mass cutoff $M_{\rm max}$ is decided by taking the upper limit of the
circular velocity $v_{\rm circ}\sim200$ km/s, beyond which the \HI content falls
off.  \citet{pontzen08} have shown  that halos more massive than 
$10^{11}$ $M_{\odot}$ do not contain a significant  amount of neutral gas.

% % % % % % % % % Figure of density fields % % % % % % % % % % % % % % % %
\begin{figure*}

%---------------------- grey figures------------------------------------	
%	\psfrag{z=6}[c][c][1.2][0]{\bf $z=6$} % for grey figures
%	\psfrag{z=3}[c][c][1.2][0]{\bf $z=3$} % for grey figures
%	\psfrag{z=1}[c][c][1.2][0]{\bf $z=1$} % for grey figures
%	\psfrag{Dark Matter}[c][c][1.8][0]{\bf Matter} % for grey figures
%	\psfrag{Halo}[c][c][1.8][0]{\bf Halo} % for grey figures
%	\psfrag{HI}[c][c][1.8][0]{\bf \HI} % for grey figures
%------------------------------------------------------------------------
	\psfrag{Mpc}[c][c][1.4][0]{Mpc}
	
%---------------------------- colour figures ----------------------------
	\psfrag{z=6}[c][c][1.2][0]{\textcolor{green}{\bf $z=6$}} % for colour figures
	\psfrag{z=3}[c][c][1.2][0]{\textcolor{green}{\bf $z=3$}} % for colour figures
	\psfrag{z=1}[c][c][1.2][0]{\textcolor{green}{\bf $z=1$}} % for colour figures
	\psfrag{Dark Matter}[c][c][1.8][0]{\textcolor{green}{\bf Matter}} % for colour figures
	\psfrag{Halo}[c][c][1.8][0]{\textcolor{green}{\bf Halo}} % for colour figures
	\psfrag{HI}[c][c][1.8][0]{\textcolor{green}{\bf \HI}} % for colour figures

%------------------------------------------------------------------------	
	
	\centering
	\includegraphics[width=1.0\textwidth, angle=-90]{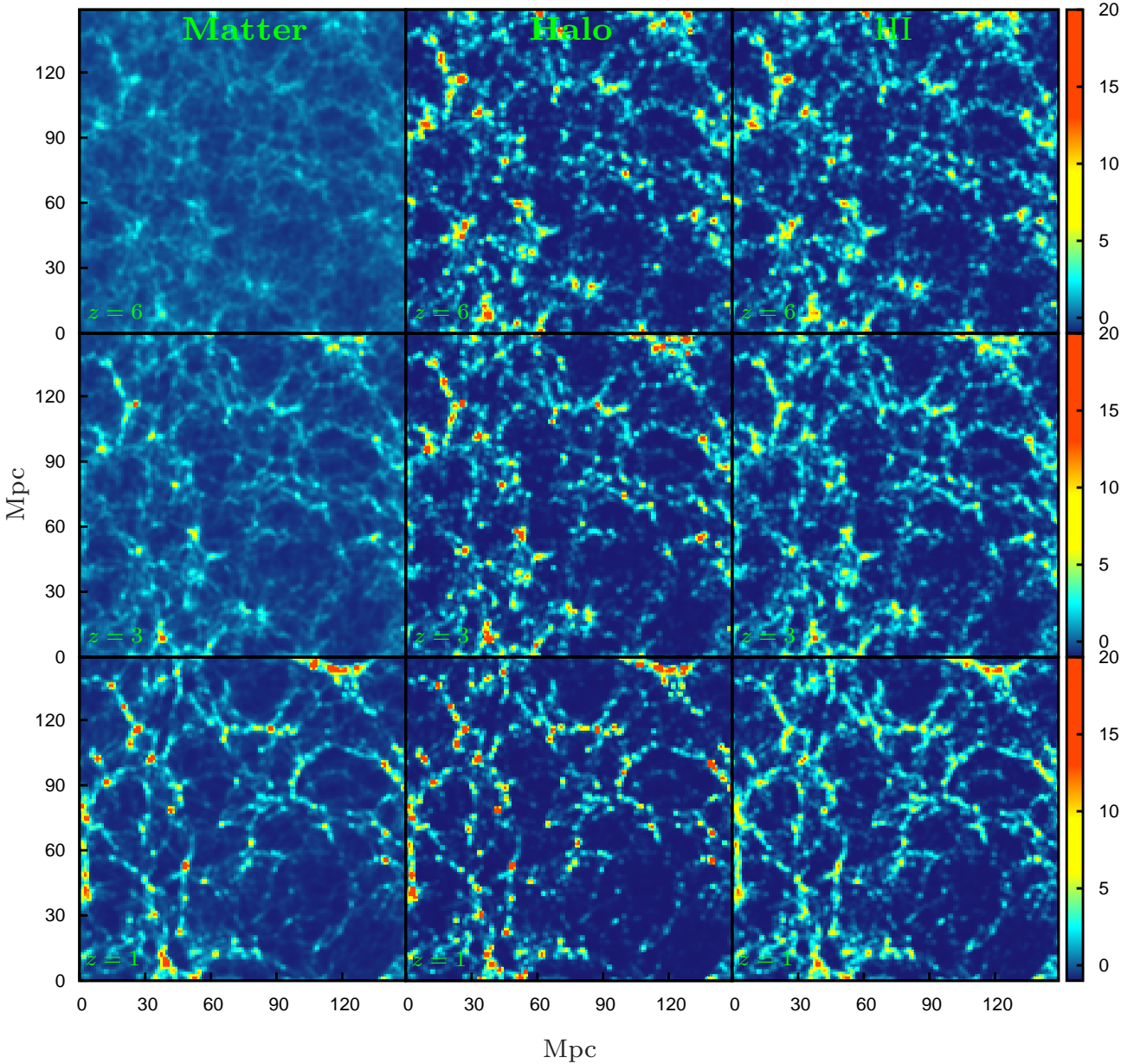}

	\caption{Shown in this figure are the Matter (left panels), Halo (central panels) and \HI (right panels) density contrasts [$\delta(\x,t)=\delta\rho(\x,t)/\bar{\rho}(t)$] respectively at three different redshifts $6,3 \, \rm{and} \, 1$ (from top to bottom). First We calculate the over densities on the grid positions using cloud in cell (CIC) interpolation scheme. We prepare the two dimensional density plots by collapsing a layer of thickness $5.6$ Mpc along one direction to calculate the average density contrast on a plane. Different colours indicate the values of density contrast on each of the pixels as shown by the colour bar.}
	\label{fig:density_dist}
\end{figure*}
% % % % % % % % % % % % % % % % % % % % % % % % % % % % % % % % % % % % % % %

In our work we have used the third scheme proposed by 
\citet{bagla10} where  the \HI mass in a halo is related to 
$M_h$ as 
\begin{equation}  
  M_{\rm \HI}(M_h) = \left\{  \begin{array}{l l}	 
    f_3\frac{M_h}{1+\left(\frac{M_h}{M_{\rm max}}\right)} & \quad
    \text{if $M_{\rm min}\leqslant M_h$} \\ 0 & \quad
    \text{otherwise}\\
  \end{array} \right.  \,.
  \label{eq:6}
\end{equation}  
According to this scheme only   halos with mass greater than  $M_{\rm min}$ 
host \HI.   The \HI mass of a halo increases proportionally 
 with the halo mass $M_{h}$ for $M_h \ll M_{\rm max}$.  However, 
 the \HI mass saturates as  $M_h \sim   M_{\rm max}$, and  $  M_{\rm \HI}$ 
attains a constant upper limit  $ M_{\rm \HI} =f_3 M_{\rm max}$ for $M_h \gg
M_{\rm max}$.   The free parameter $f_3$ determines the total amount of \HI in
the simulation 
 volume, and its value  is tuned so that it produces the desired
 value of the \HI density parameter $\Omega_{\rm \HI}\sim10^{-3}$.  Our entire
 work here deals with the dimensionless \HI density contrast $\delta \rho_{\rm
 HI}/ \bar{\rho}_{\rm HI}$ which is insensitive to the choice of $f_3$.

We have run five statistically independent realizations of the simulation.  
These  five independent realizations  were
 used to estimate the mean value and the variance for all the results
 presented in this paper.  The simulations require a large computation time, 
particularly  the FoF which takes  $\sim 10$ days for a single realization on
our computers and this restricts us to run only five independent 
realizations. The 
computation time increases at lower redshifts, and we have restricted our
simulations to $z \ge 1$. 
 
As mentioned earlier, our simulations have a halo mass resolution of  $M_h =
10^9M_{\odot}$, but eq. \ref{eq:6} shows that the mass of the smallest
possible halo that retain \HI falls as $M_{\rm min} \propto
(1+z)^{-\frac{3}{2}}$  and so, $M_{\rm min}= 10^{9} \, M_{\odot}$ at $z$=3.5,
{\it i.e.}, the minimum resolvable halo mass, $M_{\rm min}$,  falls below our mass
resolution of 10$^{9}$ M$_{\odot}$ at $z > 3.5$.  At these 
redshifts therefore, our simulations cannot detect halos less massive than
this threshold and which according to the model proposed by \citet{bagla10},
are also likely to host some \HI. 
To study the effects of these missing low mass halos  we have run  
a  high resolution simulation (referred to as HRS) with  $[2,144]^3$ particles
in a $[4,288]^3$  regular cubic grid of spacing $0.035 \, {\rm Mpc}$, the
total simulation volume remaining  the same as earlier. The lower mass limit
for the halo mass is  $10^{8.1} \, M_{\odot}$ in the HRS, well
below  $M_{\rm min}$ in the entire  redshift range.  The HRS requires
considerably larger computational resources compared to the other simulations,
and we have run only a single realization for which we have  compared
the results with those from the earlier lower resolution simulations.

\section{Results}
\label{sec:results}

Figure \ref{fig:density_dist}  provides  a visual impression of how the matter,
the halos and the \HI  are distributed at different stages of the
evolution. We show this by  plotting  the density contrasts
$\delta(\x,t)=\delta\rho(\x,t)/\bar{\rho}(t)$ 
at three different redshifts, viz. $6,\, 3 \, \rm{and} \, 1$. 
It can be seen that the cosmic web  is clearly visible in all three 
components  even  at the highest redshift $z=6$, though it is somewhat 
diffused  for the matter at this redshift. Observe that the basic skeleton of the cosmic web
is nearly the same for all the three components,   and the basic skeleton  
does not change significantly with redshift.  We see that for all the three
components the cosmic web become more prominent with decreasing redshift.   
Considering the matter first, the density contrast grows  with decreasing  redshifts
due to gravitational  clustering. 
 The halos are preferentially located at the 
matter  density peaks, and  it is evident that the halos have a higher density  
contrast. We see that the structures in the halo distribution are more
prominent  compared to the matter, particularly  at high redshifts. 
The \HI closely follows the halo distribution at $z=6$. However, 
in contrast to the matter and  halo distribution, the \HI   distribution shows a
much weaker evolution with $z$. It is possible to understand this 
 in  terms of the  model for populating the  halos with HI.
 We know that the halo masses increase as gravitational clustering proceeds. 
 According to our \HI population model,   however,  the \HI mass  remains
 fixed once the halo mass exceeds a critical value.

% % % % % % % % % % % % % % % % % %
\begin{figure*}
	\psfrag{hpk}[c][c][1.2][0]{${\Delta^{2}_{\HI}(k)}$}
	\psfrag{pkdm}[c][c][1.2][0]{${\Delta^{2}(k)}$}
	\psfrag{ck}[c][c][1.2][0]{$\Delta^{2}_{c}(k)$}
	\psfrag{k}[c][c][1.2][0]{$k\,{\rm Mpc}^{-1}$}
	\psfrag{redshift1}[c][c][1.2][0]{$z=1$}
	\psfrag{redshift2}[c][c][1.2][0]{$\quad\:\:\:$$2$}
	\psfrag{redshift3}[c][c][1.2][0]{$\quad\:\:\:$$3$}
	\psfrag{redshift4}[c][c][1.2][0]{$\quad\:\:\:$$4$}
	\psfrag{redshift5}[c][c][1.2][0]{$\quad\:\:\:$$5$}
	\psfrag{redshift6}[c][c][1.2][0]{$\quad\:\:\:$$6$}
	\psfrag{0.01}[c][c][1][0]{$10^{-2}$}
	\psfrag{}[c][c][1][0]{}
	\psfrag{}[c][c][1][0]{}

	\centering
	\includegraphics[width=0.33\linewidth, angle=-90]{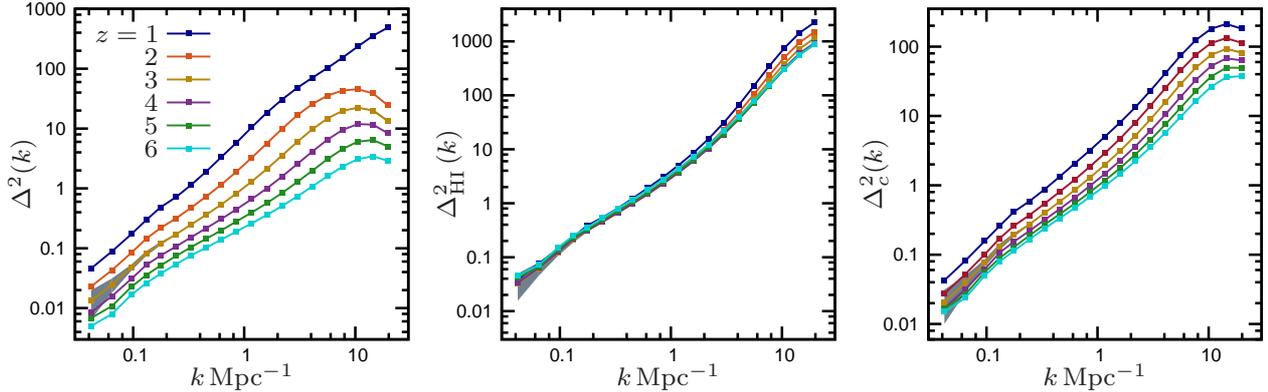}
	\caption{The dimension less form of the matter (left panel) and the \HI (central panel) power spectrum,  and the \HI-matter cross-correlation power spectrum (right panel) are shown here as a function of $k$ at six different redshifts. The shaded region in all three figures show $\pm1 \, \sigma$ error around the mean value at $z=3$.}
	\label{fig:powspecs}
\end{figure*}
% % % % % % % % % % % % % % % % % % % % % % % % %

We quantify the matter and the \HI distributions with the respective power
spectra $P(k)$ and $P_{\HI}(k)$.  We also quantify the cross correlation
between the matter and the \HI through the cross-correlation power spectrum
$P_{c}(k)$.  For all the three power spectra    we consider the 
dimensionless quantity $\Delta^{2}(k)=k^{3}P(k)/2 \pi^{2}$,
respectively shown in the three panels of Figure
\ref{fig:powspecs} for different values of the redshift $z\in[1,6]$.  
The five independent realizations of the simulation each provides a statistically 
independent estimate of the power  spectrum.  We have used these to quantify  
the mean and the standard deviation  which we show in the figures. For
clarity of presentation, the $\pm \, 1\, \sigma$ confidence interval is shown  for
$z=3$ only.

The left panel of Figure \ref{fig:powspecs} shows  $\Delta^{2}(k)$ as
a function of $k$ at different redshifts.  The matter distribution, whose
evolution is well understood (e.g. Chapter 15 of  \citealt{peacock99})  serves
as the reference against which we compare the \HI distribution at different
stages of its evolution.  
It is evident  that   $\Delta^{2}(k)$  increases  proportional to the square
of the growing mode leaving the shape of the power spectrum unchanged at small
$k$ or large  length-scales where  the predictions of linear theory 
hold (e.g. Chapter 16 of  \citealt{peacock99}).  At small scales, where
nonlinear clustering is important, the  shape of  $\Delta^{2}(k)$  changes
with  evolution and the growth  
is more than what is predicted by  linear theory. Note
  that the different power spectra shown in this paper  have all been
  calculated using a   grid whose spacing is double of that used for the
  simulations. The turn over seen in  $\Delta^{2}(k)$ at $k \,\sim \,10 \,
  {\rm  Mpc}^{-1}$ is an artefact introduced by the smoothing at this grid
  scale. We have restricted the entire analysis of this paper to the range $k
  \le 10 \, {\rm     Mpc}^{-1}$. 

 The central  panel of Figure
\ref{fig:powspecs} shows  ${\Delta^{2}_{\HI}(k)}$  as 
a function of $k$ at different redshifts. We can clearly see that the
evolution of  $\Delta^{2}(k)$ and ${\Delta^{2}_{\HI}(k)}$ are quite
different.   At small $k$,  we find that  ${\Delta^{2}_{\HI}(k)}$ shows almost 
no evolution over the entire  redshift  range. We find this behaviour over the
entire region where the matter exhibits linear gravitational clustering. 
We find that  ${\Delta^{2}_{\HI}(k)}$ grows to some extent at  
$k>2 \, {\rm  Mpc}^{-1}$ where non-linear effects are important. This growth,
however, is smaller than the growth of the matter power spectrum.  Further, we also see
that the shape of ${\Delta^{2}_{\HI}(k)}$  closely resembles
$\Delta^{2}(k)$ at small $k$, however the two differ at large
$k$,  and these differences are easily noticeable at  $k>2 \, {\rm
  Mpc}^{-1}$.

The right panel of Figure \ref{fig:powspecs} shows  $\Delta^{2}_{c}(k)$ as
a function of $k$ at different redshifts. We see that the evolution of 
$\Delta^{2}_{c}(k)$ is intermediate to that of  $\Delta^{2}(k)$ and 
${\Delta^{2}_{\HI}(k)}$, it grows faster than  ${\Delta^{2}_{\HI}(k)}$ but not
as fast as  $\Delta^{2}(k)$.  All three power spectra have the same shape
at small $k$, indicating that the \HI traces the matter at large
length-scales.  At large $k$ the shape of $\Delta^{2}_{c}(k)$, however, differs from both 
 $\Delta^{2}(k)$ and  ${\Delta^{2}_{\HI}(k)}$ indicating differences in
the small-scale clustering of the \HI and the matter.

Redshift surveys of large scale structures and numerical simulations  reveal
that the galaxies trace underlying matter over-densities with a possible bias
\citep{bbks-bardeen-bond-kaiser-szalay86,mo-white96,dekel-lahav99}. In the post-reionization era, the  
association of the \HI with the halos implies that the \HI  follows the matter
density field with some bias. The bias function relates the  \HI density contrast
to that of the matter.
Here we assume  that a linear relation holds between the Fourier components of the
\HI and the matter density contrasts
\begin{equation}
 \Delta_{\HI}(\k) = \tilde{b}(\k)\, \Delta(\k)
 \label{eq:7}
\end{equation}
where, $\tilde{b}(\k)$ is the linear bias function or simply bias, which can, 
in general, be complex. The complex bias allows for the
  possibility that the Fourier modes  $\Delta_{\HI}(\k)$ and $ \Delta(\k)$
  can  differ in both the  amplitude and  also the phase. 
The ratio of the respective power spectra 
\begin{equation}
b(k)=\sqrt{\frac{P_{\HI}(k)}{P(k)}} \,.
\label{eq:8}
\end{equation}
allows us to quantify  $b(k)$ which is the modulus  of the complex bias 
$\tilde{b}(k)$, and the ratio 
\begin{equation}
b_r(k)=\frac{P_{c}(k)}{P(k)} \,.
\label{eq:9}
\end{equation}
allows us to quantify  $b_r(k)$ which is the real part  of the complex bias  
$\tilde{b}(k)$. With both $b(k)$ and $b_r(k)$ at hand, we can reconstruct the
entire complex bias $\tilde{b}(k)$. One is mainly  interested in the modulus 
 $b(k)$ which allows us  to interpret the \HI
power spectrum in terms of the underlying matter power spectrum.  However, the
real part of the bias $b_r(k)$ is the relevant quantity if one is considering
the cross-correlation of the \HI with either the matter distribution
or with some other tracer of the matter distribution like Lyman-$\alpha$ forest 
\citep{tapomoy-somnath-trc-kanan11,tapomoy-kanan15} or galaxy surveys \citep{chang-pen-bandura10,masui-switzer-banavar13,cohn-white-chang-holder15}.

\begin{figure*}
\psfrag{bkmod}[c][c][1.2][0]{$b(k)$}
\psfrag{bbcr}[c][c][1.2][0]{$b(k)$ and $b_{r}(k)$}
\psfrag{bz}[c][c][1.2][0]{$b(z)$} 
\psfrag{bias}[c][c][1][0]{$b(k)\:\:\:$} 
\psfrag{crbias}[c][c][1][0]{$b_r(k)$} 
\psfrag{z}[c][c][1.2][0]{$z$}
\psfrag{k}[c][c][1.2][0]{$k\, {\rm Mpc}^{-1}$}
\psfrag{redshift5}[c][c][1.2][0]{$\quad\:\:\:$$5$} 
\psfrag{redshift6}[c][c][1.2][0]{$\quad\:\:\:$$6$} 
\psfrag{rsvalueis4}[c][c][1.2][0]{$\:$$z=4$} 
\psfrag{percentage difference}[c][c][1][0]{Fractional difference in $\%$}

\psfrag{redzz1}[c][c][1.2][0]{$z=1$}
\psfrag{redzz2}[c][c][1.2][0]{$\:\:\:\quad2$}
\psfrag{redzz3}[c][c][1.2][0]{$\:\:\:\quad3$}
\psfrag{redzz4}[c][c][1.2][0]{$\:\:\:\quad4$}
\psfrag{redzz5}[c][c][1.2][0]{$\:\:\:\quad5$}
\psfrag{redzz6}[c][c][1.2][0]{$\:\:\:\quad6$}
	
\centering \includegraphics[width=.33\linewidth,
                  angle=-90]{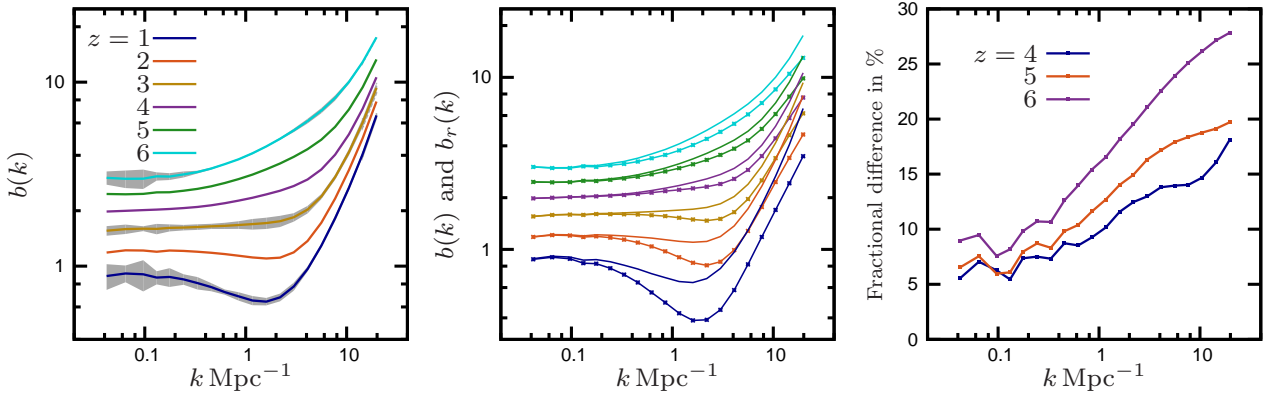}
\label{fig:bias}
\caption{The left panel shows the $k$ dependence of the \HI bias $b(k)$ at six
different redshifts, $z=6 \, - \, 1$ (top to bottom) at an interval $\Delta
z=1$. The shaded regions show $\pm 5 \, \sigma$ error around the mean value
for three redshifts $6$,$3$ and $1$. 
The central panel shows the $k$ dependence of both the biases, $b(k)$ (line
only)  and $b_{r}(k)$ (line-point), at six different redshifts $z=6 \, - \, 1$
(top to bottom) at an  interval $\Delta z=1$. For $z \ge 4$,  
the right panel quantifies the effect  of a fixed minimum halo mass which has
a value  $M_{\rm min} = 10^{9}M_{\odot}$ in the low resolution
simulations. The figure   shows the fraction difference in  the  bias  $b(k)$ relative to a 
high resolution simulation (HRS). }
\end{figure*}

The left panel of Figure \ref{fig:bias} shows the behaviour of $b(k)$, 
the modulus  of $\tilde{b}(k)$,  as a
function of $k$ at six different redshifts. We also show the  $5 \, \sigma$
confidence interval at three different redshifts. The relatively small  errors 
indicate that the results  reported  here are statistically representative 
values. We see 
that the value of $b(k)$ decreases  with decreasing redshift. Further, the $k$ dependence
is also seen to evolve with redshift. In all cases, we have a constant  
$k$ independent bias  at small $k$ and the bias shoots up rapidly with $k$ at
large $k$  ($\ge 4 \, {\rm Mpc}^{-1}$). However, for high redshifts $(z \ge
3)$ , $b(k)$ increases monotonically with $k$ whereas we see a   dip in the
values of $b(k)$ at $k \sim  2\, {\rm Mpc}^{-1}$ for $z < 3$. Interestingly,
the $k$ range where we have  a constant $k$ independent bias   is maximum at
the intermediate redshift $z = 3$ where it  extends to $k \le 1 \, {\rm
  Mpc}^{-1}$, and it is minimum  ($k \le 0.2 \, {\rm  Mpc}^{-1}$)
at the highest and lowest redshifts  ($z=6,1$)  whereas it covers 
 $k \le 0.4 \, {\rm  Mpc}^{-1}$ at the other redshifts $(z=2,4,5)$.

The central panel of Figure \ref{fig:bias} shows  both $b(k)$ and  
$b_r(k)$  which is the real part   of $\tilde{b}(k)$. 
The two quantities  $b(k)$ and  $b_r(k)$ show similar $k$
dependence.
Both  $b(k)$ and  $b_r(k)$ will be equal 
if the bias $\tilde{b}({k})$ is a real quantity.
We see that this is true at small $k$  where both quantities 
have nearly constant values independent of $k$. 
However, we find  a $k$ independent bias $b_r(k)$ for a smaller 
range of $k$, in comparison to $b(k)$.
The two quantities  $b(k)$ and  $b_r(k)$  differ at larger $k$, 
and the differences increase with increasing $k$.  
The difference  between $b(k)$  and  $b_r(k)$ is seen to increase with
decreasing redshift. Also the $k$ value where these differences become
significant shifts to smaller $k$  with decreasing redshift. 

\begin{figure*}
	
	\psfrag{bias and real bias}[c][c][1.2][0]{$b(k)$ and
          $b_{r}(k)$} 
    \psfrag{z}[c][c][1.2][0]{$z$} 
    \psfrag{k1 value}[c][c][1][0]{$k=0.065\, \rm{Mpc}^{-1}$} 
    \psfrag{ksmall}[c][c][0.8][0]{$k=0.001\, \rm{Mpc}^{-1}$$\qquad\qquad\qquad\qquad\qquad$} 
    \psfrag{k2 value}[c][c][1][0]{$k=0.45\, \rm{Mpc}^{-1}$} 
    \psfrag{k3 value}[c][c][1][0]{$k=2.2\, \rm{Mpc}^{-1}$} 
    \psfrag{fit to the data of bias}[c][c][0.8][0]{$b(k)$ fitting$\quad\quad\;$} 
    \psfrag{fit to the real part}[c][c][0.8][0]{$b_{r}(k)$ fitting$\quad\quad\quad$}
    \psfrag{bias data points only}[c][c][0.8][0]{$b(k)$ simulations}
    \psfrag{real part data points}[c][c][0.8][0]{$b_{r}(k)$ simulations}
    
    \centering \includegraphics[width=0.37\linewidth,
      angle=-90]{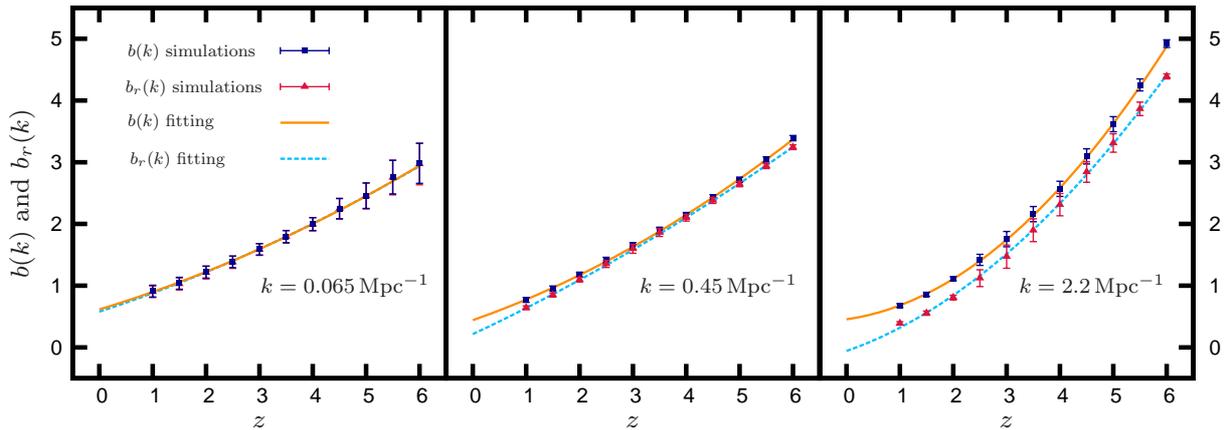}
    \caption{This shows the redshift evolution of $b(k)$
      and $b_{r}(k)$ at three different $k$ values. The  points 
and the vertical error bars respectively show the mean and $5 \sigma$
spread determined from $5$ realisations of the simulations. The solid and
dotted lines show   the fitting of the respective quantities.}
	\label{fig:diff_bias}
\end{figure*}

As already mentioned in Section \ref{sec:simulations}, for $z>3.5$,  $M_{\rm
  min}$ (eq. \ref{eq:5}) has a value that is smaller than $10^{9} \, M_{\odot}$ 
which is the smallest mass halo resolved by our simulations.  Imposing a fixed
lower halo mass limit of $10^{9} \, M_{\odot}$ will, in principle, change the
\HI bias $\tilde{b}(k)$ in comparison to the actual predictions of the halo
population model proposed by \citet{bagla10}, and we have run higher
resolution simulation in order to quantify this.  It is
computationally expensive to run several realizations of  simulations with a
smaller mass resolution, so we have run a single realization 
with a halo  mass resolution of   $10^{8.1} \, M_{\odot}$ which is well below 
 $M_{\rm  min}$ over the entire redshift range of our interest. 
The right  panel of Figure \ref{fig:bias} shows   the percentage difference  in
the values of $b(k)$ computed using the low and the high resolution
simulations.  We find that   the difference is minimum for $z=4$ and
maximum for $z=6$  where we expect a larger contribution from the  smaller
halos.  For $k<1.0\, \rm{Mpc}^{-1}$, the difference is $5 \, - \, 8\%$ at $z=4$ and $8 \, - \, 13\%$ at
$z=6$.  Beyond $1.0\, \rm{Mpc}^{-1}$, the difference increases but it is well
within $20\%$ for redshifts $4\, \rm{and} \,5$ and less than $30 \%$ for
redshift $6$. These differences are relatively small given our current lack of
knowledge about how the \HI is distributed at the redshifts of interest. It is
therefore well justified to use the simulations with a fixed lower mass limit
of $10^{9} \, M_{\odot}$ for the entire redshift range considered in this
paper.

Figure \ref{fig:diff_bias} shows the redshift evolution of $b(z)$ and
$b_{r}(z)$ at three representative  $k$ values. 
At $k=0.065 \, {\rm Mpc}^{-1}$ (left panel) which is in the linear regime we
cannot make out the difference between $b(z)$ and $b_{r}(z)$ and this  indicates
that $\tilde{b}(z)$ is purely real. We find  that the bias $b(z)$ has a value
that is slightly less than unity at $z=1$ indicating that the
\HI is slightly anti-biased  at this redshift. The bias increases nearly
linearly with  $z$ and it has a value $b(z) \approx 3$ at $z=6$.  At $k=0.45
\, {\rm Mpc}^{-1}$ 
(central panel) where we have the transition from  the linear to  the
non-linear regime  we find that $b(z)$ is slightly larger than $b_r(z)$. 
Both $b(z)$ and $b_r(z)$ show a $z$ dependence very similar to that in the
linear regime. At $k=2.2 \, {\rm Mpc}^{-1}$ (right panel) which is in the non-linear
regime we find that $b_{r}(z)$ is  appreciably smaller than  $b(z)$, and the
difference is nearly constant over the entire $z$ range. This indicates that
the \HI bias $\tilde{b}(z)$ is complex in the non-linear regime. Further, we
see that the relative contribution from the imaginary part of  $\tilde{b}(z)$
increases  with decreasing $z$. The value of 
 $b_{r}(z)$ is less than unity for $z \le 2$, whereas this is so only in the
range   $z \le 1.5$ for $b(z)$. The redshift dependence of the bias is much  
steeper as compared to the linear regime,  and we have a larger value of $b(z)
\approx 5$ at $z=6$. We find a nearly parabolic $z$ dependence in the
no-linear regime as  compared to the approximately linear redshift dependence
found at smaller $k$.  At all the three $k$ values     we have fitted the
redshift evolution of $b(z)$ and $b_{r}(z)$ with a  quadratic polynomial 
of the form  $b_{0} + b_{1}z + b_{2}z^{2}$. 
We find that the polynomials give a very good fit to the redshift evolution of
the simulated data (Figure \ref{fig:diff_bias}).  We also find that the
quadratic term $b_2$ is much larger at $k=2.2 \, {\rm Mpc}^{-1}$  as compared
to   the two smaller $k$ values. Based on these results, we have carried out a
joint fitting of the $k$ and $z$ dependence of the bias, the details of which
are presented in the next section.

\subsection{Fitting the  bias}
\label{subsec:fitting}

% % % % % % % % % % % % % % % % % % % % % % % % % % % % % % % % %
\begin{figure*}
  
  \psfrag{bkmod}[c][c][1.2][0]{$b(k)$} 
  \psfrag{bkreal}[c][c][1.2][0]{$b_{r}(k)$} 
  \psfrag{k}[c][c][1.2][0]{$k\, {\rm Mpc}^{-1}$} 
  \psfrag{b1}[c][c][1.2][-90]{$b_1\quad \quad$} 
  \psfrag{b0}[c][c][1.2][-90]{$b_0\quad \quad$} 
  \psfrag{b2}[c][c][1.2][-90]{$b_2\quad \;$} 
  \psfrag{b3}[c][c][1.2][-90]{$b_3\quad \;$}
  \psfrag{b4}[c][c][1.2][-90]{$b_4\quad$} 
  \psfrag{z}[c][c][1.2][0]{$z$}
  \psfrag{bmo}[c][c][0.7][0]{$b_m\:\:$}
  \psfrag{bmr0}[c][c][0.7][0]{$b_{m0}$}
  \psfrag{}[c][c][1][0]{}
  \psfrag{}[c][c][1][0]{}
  \psfrag{}[c][c][1][0]{}
  \psfrag{}[c][c][1][0]{}
  
  \centering \includegraphics[width=0.33\linewidth,
    angle=-90]{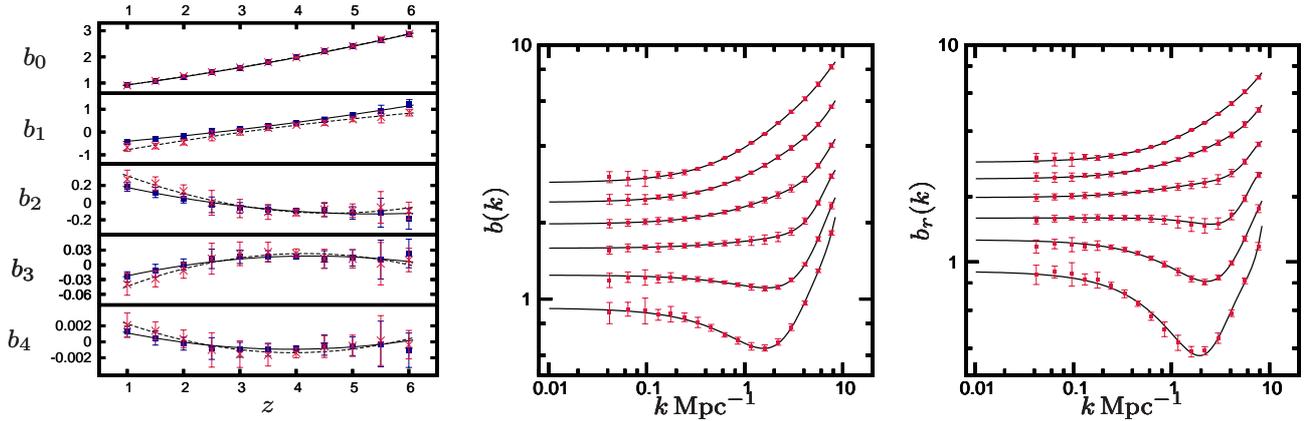}
  \caption{The left panel shows the redshift evolution of the coefficients of fitting,  $b_m$ and $b_{rm}$. The square points represent the $b_m$ values and the cross points represent the $b_{rm}$ values. The errors in fitting are enhanced $5$ times to make them visible and shown with the vertical error bars. The solid and the dotted black lines show the best fit curves for $b_m$ and $b_{rm}$ respectively. The $k$ dependence of the simulated biases, $b(k)$ and $b_r(k)$, at six different redshifts  $z=1-6$ (bottom to top) at an interval $\Delta z=1$, are respectively shown in the central and in the right panel with $5\, \sigma$ error bars. The black dotted lines show the best fit curves.}
  \label{fig:fit_bias}
\end{figure*}
% % % % % % % % % % % % % % % % % % % % % % % % % % % % % % % %

We have carried out polynomial fitting for  the $k$  dependence of the bias
(Figure \ref{fig:bias}). The fit was  carried out for redshifts in the
range $z=1$ to $6$ at an interval  of  $\Delta z=0.5$.  
 We find that a linear function   of the form $b(k) =
b_0 + b_1 k$ gives a good fit to the simulated data for $z \ge 4$. However,  a
higher order  polynomial is required at lower redshifts particularly because
of the dip around $k \sim 2 \, {\rm Mpc}^{-1}$.   We have used  a  quartic
polynomial of the form  
\begin{equation}
b(k)=b_0 + b_1 k + b_2 k^2 + b_3 k^3 + b_4 k^4\,.
\label{eqn:8}
\end{equation}
which gives a good fit in  the $k$~range  $k \leq 10{\rm Mpc}^{-1}$ at all the
redshifts that we have considered.  
The fit was carried out for both $b(k)$ and $b_r(k)$, and we have retained  the 
subscript $r$ for the different fitting coefficients of  $b_r(k)$.

The left panel of Figure \ref{fig:fit_bias} shows how the $5$ fitting
coefficients $b_0,...,b_4$ vary with redshift. The value of the coefficient
$b_0$ corresponds to the  scale independent bias which is seen to hold at
small $k$ values. We find that $b_0$ and $b_{r0}$ are indistinguishable over
the entire redshift range, indicating that the bias is real at small  
 $k$ values. We also find that $b_0$ increases nearly linearly with $z$,
consistent with the behaviour seen in the left panel of Figure
\ref{fig:diff_bias}. The coefficients $b_1,...,b_4$ introduce a scale 
dependence in the bias, and these coefficients have progressively smaller
values. We find that the redshift dependence of all the five coefficients can
be well fit by quadratic polynomials of the form 
\begin{equation}
b_m(z)=c(m,0)+c(m,1) z + c(m,2) z^2,
\label{eq:fit}
\end{equation}
the fits also being shown in the left panel of Figure \ref{fig:fit_bias}. 
The fitting coefficients $c(m,n)$  allow us to interpolate  
the bias $b(k,z)$  at different values of $z$ and $k$ in the ranges $[1,6]$
and $[0.04,10]$ respectively. The fitting coefficients $c(m,n)$ and the
$1 \, \sigma$ errors in these coefficients $\Delta c(m,n)$ are tabulated in the
Appendix \ref{sec:appendix}.  The central panel of Figure \ref{fig:fit_bias}   
shows the fit along with  the simulated data. We see that the fit reproduces 
the simulated data to a good level of accuracy over the entire $z$ and $k$
range of the fit.  A similar fitting procedure was also carried out for
$b_r$. The fitting coefficients $c_r(m,n)$ and the
$1\, \sigma$ errors in these coefficients $\Delta c_r(m,n)$ are tabulated in the 
Appendix \ref{sec:appendix}. The right panel of Figure \ref{fig:fit_bias} shows that the fit
matches the simulated $b_r$ values to a good level of accuracy. 

% % % % %
\begin{figure}
	
  \psfrag{kk}[c][c][1.2][0]{$k\, {\rm Mpc}^{-1}$} 
  \psfrag{z}[c][c][1.2][0]{$z$}
  \psfrag{bk}[c][c][1.3][0]{$b(k,z)$}
  \psfrag{brk}[c][c][1.3][0]{$b_r(k,z)$}
  \psfrag{}[c][c][1][0]{}
  \psfrag{}[c][c][1][0]{}
  
  \centering \includegraphics[width=0.6\linewidth,
    angle=-90]{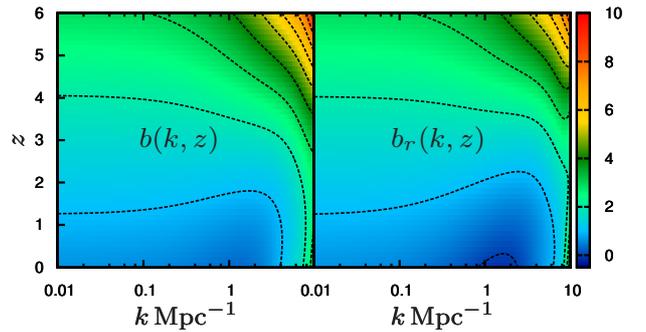}
  \caption{The joint $k$ and $z$ dependence of the biases $b(k,z)$ (left panel) and $b_r(k,z)$ (right panel) are shown here. The values of $b(k,z)$ and $b_r(k,z)$ at different points of $k-z$ plane is represented with appropriate colours. The contours are drawn through those $k$ and $z$ values where the biases $b(k,z)$ and $b_r(k,z)$ have values in the range $0 \, - \, 10$ (bottom to top) at an interval of $1$.}
  \label{fig:cont}
\end{figure}
% % % % % % % % % % %

Figure \ref{fig:cont} provides a visual impression of how the bias varies
jointly with $k$ and $z$. Here we have extrapolated our fit to cover a somewhat
larger $k$ range ($[0.01,10] \, {\rm Mpc}^{-1}$) and $z$ range ($[0,6]$).  We find a scale
independent bias for $k \le 0.1 \, {\rm Mpc}^{-1} $ across the entire $z$
range. Further, we  see that the
biases $b(k,z)$ and $b_r(k,z)$ both decrease monotonically with decreasing
$z$. We also see that the \HI and the matter become anti-correlated 
where $b_r$ has a negative value 
for   the $k$ range  $k \sim  1-2 \, {\rm Mpc}^{-1}$ around $z \sim 0$. 

% % % % % % % % % % % % % % % % % % %

\begin{figure}
	
	\psfrag{kk}[c][c][1.2][0]{$k\, {\rm Mpc}^{-1}$} 
	\psfrag{z}[c][c][1.2][0]{$z$}
	\psfrag{}[c][c][1][0]{}
	\psfrag{}[c][c][1][0]{}
	
	\centering \includegraphics[width=0.45\textwidth,
	angle=-90]{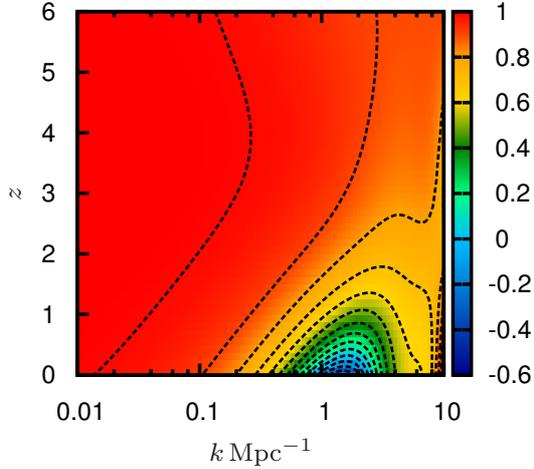}
	\caption{The joint $k$ and $z$-dependence of stochasticity parameter
          $r(k,z)\,\equiv b_{r}(k,z)/b(k,z)$ is shown here. The contours are
          drawn for the value of $r=1$ to $-0.5$ at an
          interval $\Delta r=0.1$ (left to right). The entire region to the
          left of the $r=1$ contour corresponds to a fixed value of $r=1$.}
	\label{fig:stocpar}
\end{figure}

The cross-correlation between the  \HI and the matter can also be
quantified using the stochasticity \citep{dekel-lahav99}  $r=b_r/b$.
By definition $\mid r \mid \leq 1$, values $r \sim 1$ indicate a strong correlation,
$r \sim 0$ corresponds to a situation when the two are uncorrelated and $r <
0$ indicates anti-correlation.  Figure \ref{fig:stocpar} shows
how the stochasticity $r$ varies jointly as a function of $z$ and $k$. We 
see that $r=1$ for $k \le 0.1 \, {\rm Mpc}^{-1} $ 
where the bias also is   scale independent and real across the
entire $z$ range.  The $k$ value below
which $r$ is unity increases with increasing redshift, with 
$k \sim 0.1 \, {\rm Mpc}^{-1} $ for $z=0$ and $k \sim 3 \, {\rm Mpc}^{-1} $
for $z=6$. The \HI and the matter are highly correlated $(r > 0.8)$
across nearly the entire $k$ range for $z \ge 2$.  We also find that $r$ has
a negative value at $k \sim 1 \, - \, 2\, {\rm Mpc}^{-1}$ around $z \sim 0$,
indicative of an anti-correlation.

\section{Summary and Discussion}
\label{sec:summary}

In this paper we have used semi-numerical simulations
(the third scheme of \citet{bagla10}) to model the \HI distribution 
and study the evolution of $P_{\HI}(k,z)$ in   the post-reionization  era.   The
simulations span the  redshift range $1 \le z \le 6$   at an interval $\Delta
z =0.5$.   We have modelled the \HI bias  as a complex quantity
$\tilde{b}(k,z)$  whose modulus  $b(k,z)$ (squared) relates $P_{\HI}(k,z)$ to
$P(k,z)$,  and whose real part   $b_r(k,z)$ quantifies the cross-correlation
between the \HI and the total matter distribution.
While there are several earlier works which have  studied the \HI bias $b(k,z)$
at a few discrete redshifts (summarized in
\citet{hamsa-tirthankar-refregier15}), this is the first
attempt to model the post-reionization \HI  distribution across a large $z$ and
$k$ range $(0.04 \le k \le 20 \, {\rm  Mpc}^{-1})$
using a single simulation technique. 
  
We find that the assumption of a scale-independent bias $b(k,z)=b_0(z)$ holds at
small $k$  (eq. \ref{eqn:8}). The value of $b_0(z)$  increases nearly
linearly with  $z$, with a value that is slightly less than unity at $z=1$ 
and  $b_0(z) \approx 3$ at $z=6$. 
The $k$ range where we have  a constant $k$ independent bias   is maximum at
the intermediate redshift $z = 3$ where it  extends to $k \le 1 \, {\rm
  Mpc}^{-1}$, and it is minimum  ($k \le 0.2 \, {\rm  Mpc}^{-1}$)
at the highest and lowest redshifts  ($z=6,1$)  whereas it covers 
 $k \le 0.4 \, {\rm  Mpc}^{-1}$ at the other redshifts. The  bias is scale
dependent  at larger $k$ values where non-linear effects become important.   
We  find that a polynomial fit (eq. \ref{eq:fit}) provides a good description
of  the joint $z$ and $k$ dependence of $b(k,z)$ (and also $b_r(k,z)$).   
The coefficients of the fit are presented in Appendix \ref{sec:appendix}, 
and Fig. \ref{fig:cont}  provides a comprehensive picture of the 
bias across the entire $k$ and $z$ range, all the way to $z=0$ where the
results have been extrapolated from the fit.   

Our results which are based on a PM N-body code 
are qualitatively consistent with the earlier work of 
\citet{bagla10} who have used a high resolution Tree-PM N-body code 
to  calculate  the bias at three different redshifts
($z=1.3, 3.4$ and $5.1$).  The present work is also consistent with 
 \citet{tapomoy-mitra-majumder12} who have used a technique similar to ours to
 compute the bias across $z = 1.5 \, - \, 4$, and
 \citet{hamsa-tirthankar-refregier15} who have applied  the minimum variance
 interpolation technique to the different bias values collated from literature
 to predict the redshift evolution of the scale independent bias in the range
 $z = 0 \,  -  \, 3.4$.

The analytic model of  \citet{marin-gnedin10} predicts the   \HI distribution
to be  anti-biased at low redshifts $(z\, \le \, 1)$.  They also  found that the 
 bias  decreases further for  $k \ge  0.1 {\rm Mpc}^{-1}$. These predictions
 are consistent with observations at  $z\, \sim \,0.06$ \citep{martin-giovanelli12}
 which  suggest that \HI rich galaxies are very 
weakly clustered and mildly anti-biased at large scales, but become
severely anti-biased on smaller scales. The predictions of our
simulations which are restricted to $z \ge 1$  are consistent with the
findings of   \citet{marin-gnedin10}. We find that the \HI is mildly
anti-biased at large scales at $z = 1$, and the   bias drops further for  $k
\ge  0.1 {\rm   Mpc}^{-1}$ (Fig \ref{fig:bias}). We have also extrapolated our
results to $z \sim 0$ (Fig \ref{fig:cont})  where the predictions are found to
be qualitatively consistent with the measurements  of
\citet{martin-giovanelli12}.

In our analysis the real part $b_r(k,z)$ of the complex bias $\tilde{b}(k,z)$
quantifies the cross-correlation between the \HI and the total matter, and
the  bias $\tilde{b}(k,z)$ is completely real if the two  
are perfectly correlated. The same issue is also  quantified using the
stochasticity $r=b_r(k,z)/b(k,z)$. We see that $b_r$ closely matches $b$
at small $k$ ($<0.1 \,{\rm Mpc}^{-1}$) where we have a scale independent bias 
across   the entire $z$ range. The complex nature of the bias becomes
important at larger $k$. Our results are summarized in Fig. \ref{fig:stocpar}
which shows $r$ across the entire $z$ and $k$ range. We find that 
the \HI and the  matter are well correlated $(r > 0.8)$
across nearly the entire $k$ range for $z \ge 2$. 
 We also find that $r$ has a negative value at $k \sim 1 \, - \, 2\, {\rm Mpc}^{-1}$
 around $z \sim 0$, indicative of an anti-correlation.

The measurements of \citet{chang-pen-bandura10} constrain  the product
$\Omega_{\HI} \,b \,r=(5.5 \pm 1.5) \times 10^{-4}$ at $z\sim 0.8$. From our
analysis, we find that on large scales the product $b \,r\equiv b_r=0.79$ at
$z=0.8$ which implies  $\Omega_{\HI}=(6.96 \pm 1.89) \times
10^{-4}$. Again, \citet{masui-switzer-banavar13}  constrain the 
product $\Omega_{\HI}\, b\, r=(4.3 \pm 1.1) \times 10^{-4}$ at $z\sim 0.8$ 
using measurements in the $k$ range $0.05 \,{\rm Mpc}^{-1}<k<0.8 \,{\rm 
  Mpc}^{-1}$ where our work predicts  $br$ to vary from $0.83$ to
$0.39$. The 
corresponding $\Omega_{\HI}$ varies between $(5.2 \pm 1.3) \times 10^{-4}$ to
$(1.1 \pm 0.33) \times 10^{-3}$, which is a significant variation. On the 
other hand, \citet{switzer-masui-bandura-calin13} constrain  the
product $\Omega_{\HI}\, b=6.2^{+2.4}_{-1.5}\times10^{-4} $ at $z\sim 0.8$ which
implies  $\Omega_{\HI}= 7.5^{+2.9}_{-1.8}\times10^{-4}$ 
if we consider  $b=0.83$ from our analysis. 
The above estimates of $\Omega_{\HI}$ are
consistent with the measurement $\Omega_{\HI}=7.41 \pm 2.71 \times  10^{-4}$
at $z\,\sim \, 0.609$ \citep{rao-turnshek06}. We note that our simulations are
restricted to $z \ge 1$, and the results were extrapolated to $z=0.8$ 
for the discussion presented in this paragraph. 
\citet{khandai-sethi-dimatteo11} have carried out simulations which were
specifically designed to interpret the results of
\citet{chang-pen-bandura10}, and they have predicted $b=0.55\ -\ 0.65$ and   
$r=0.9 \ - \ 0.95$  at $z=0.8$. We note that the bias  value predicted by 
\citet{khandai-sethi-dimatteo11} is considerably smaller than our prediction,
and they predict $\Omega_{\HI}=11.2 \pm 3.0 \times  10^{-4}$ which also is
larger than the measurements of \citet{rao-turnshek06}.

We finally reiterate that it is important to model the \HI distribution in
order to correctly predict the signal for upcoming 21-cm intensity mapping
experiments. Further, such modelling is also important to correctly interpret
the outcome of the future observations. In the present work we have
implemented a simple \HI population scheme which incorporates the salient
features of our present understanding  {\it ie.} the \HI resides in halos 
which  also host the galaxies. This however  ignores various
complicated astrophysical processes which could possibly play a role  in
shaping the \HI distribution.  Further, the entire analysis has been
restricted to real space, and the effects of redshift space distortion have not
been taken into account. We plan to address these issues in future work. 

%----------------------------- Acknowledgement ----------------------------------------------

\section*{Acknowledgement} 

Debanjan Sarkar wants to thank Rajesh Mondal for his help with simulations.
Anathpindika, S., acknowledges support from the grant YSS/2014/000304 of the SERB, Department of Science \& Technology, Government of India.
The authors  are grateful to J. S. Bagla, Nishikanta Khandai, Tapomoy Guha
Sarkar and Kanan K. Datta for  useful discussions. 
% % % % % % % % % % % % % % % % % % % % Bibliography % % % % % % % % % % % % % % % % % % % % % %

%\bibliography{reference}

%--------------------------------------------------------------------------------------------
% % % % % % % % % % % % % % % % % % % % % End % % % % % % % % % % % % % % % % % % % % % % % %
%\section{Appendix}
\appendix
\section{}
\label{sec:appendix}

We have fitted the joint $k$ and $z$ dependence of the biases $b(k,z)$ and $b_r(k,z)$ using polynomial of the form
\begin{equation}
	b(k,z)=\sum\limits_{m=0}^{4} \sum\limits_{n=0}^{2} c(m,n) k^m z^n\,{\rm and}
	\label{eq:10}
\end{equation}
%-------------------------------------------------------------------------------
\begin{equation}
	b_r(k,z)=\sum\limits_{m=0}^{4} \sum\limits_{n=0}^{2} c_r(m,n) k^m z^n\,
	\label{eq:11}
\end{equation}

The best fit values of the fitting coefficients $c(m,n)$ and $c_r(m,n)$, and the $1 \, \sigma$ uncertainties in fitting $\Delta c(m,n)$ and $\Delta c_r(m,n)$ respectively, are given below.

%=========================================================================
$c(m,n)\times10^{-2}$=\bordermatrix{~ & $n=0$ & $1$ & $2$ \cr
		$m=0$ & 65.31 & 25.19 & 1.963 \cr
		$\quad\: 1$ & -60.74  & 18.56 & 1.806 \cr
		$\quad\: 2$ & 33.54 & -17.38 & 1.618 \cr
		$\quad\: 3$ & -5.129 & 3.247 & -0.3803 \cr
		$\quad\: 4$ & 0.2773 & -0.1899 & 0.02435\cr}

\vspace{1cm}

$\Delta c(m,n)\times10^{-3}$=\bordermatrix{~ & $n=0$ & $1$ & $2$ \cr
		$m=0$ & 16.67 & 10.45 & 1.626 \cr
		$\quad\: 1$ & 46.34 & 31.15 & 4.995 \cr
		$\quad\: 2$ & 26.0 & 18.37 & 3.015 \cr
		$\quad\: 3$ & 5.841 & 4.255 & 0.7093 \cr
		$\quad\: 4$ & 0.3850 & 0.2856 & 0.04811\cr}

\vspace{1cm}

$c_r(m,n)\times10^{-2}$=\bordermatrix{~ & $n=0$ & $1$ & $2$ \cr
		$m=0$ & 65.49 & 25.55 & 1.934 \cr
		$\quad\: 1$ & -121.5  & 45.73  & -1.952 \cr
		$\quad\: 2$ & 58.58 & -30.59 & 3.304 \cr
		$\quad\: 3$ & -9.325 & 5.597 & -0.6735 \cr
		$\quad\: 4$ & 0.5119 & -0.3273 & 0.04138\cr}

\vspace{1cm}

$\Delta c_r(m,n)\times10^{-3}$=\bordermatrix{~ & $n=0$ & $1$ & $2$ \cr
		$m=0$ & 27.49 & 15.87 & 2.274 \cr
		$\quad\: 1$ & 81.87 & 49.21 & 7.097 \cr
		$\quad\: 2$ & 51.31 & 32.15 & 4.714 \cr
		$\quad\: 3$ & 10.32 & 6.641 & 0.9892 \cr
		$\quad\: 4$ & 0.6510 & 0.4253 & 0.06411\cr}

%----------------------------------------------------------------

% % % % % % % % % % % % % % % % % % % % % % % % % % % % % % % % % % % % % % % % % % % % % %

\end{document}